\documentclass[%
 reprint,
 amsmath,amssymb,
 aps,
pra,
]{revtex4-2}

\usepackage{xcolor}
\usepackage{graphicx}
\usepackage{dcolumn}
\usepackage{bm}
\begin{document}

\title{Rayleigh-Bénard convection with phase change close to the critical point}
\author{Valentin Mouet$^1$}
\author{Guillaume Michel$^2$}
\email{guillaume.michel@sorbonne-universite.fr}
\author{Fran\c cois P\'etr\'elis$^1$}
\author{Stephan Fauve$^1$}
\affiliation{
$^1$ Laboratoire de Physique de l’\'Ecole Normale Sup\'erieure, ENS, Universit\'e PSL, CNRS, Sorbonne Universit\'e, Universit\'e de Paris - F-75005 Paris, France\\
$^2$ Sorbonne Universit\'e, CNRS, Institut Jean Le Rond d’Alembert - F-75005 Paris, France
}

\date{\today}

\begin{abstract}
Rayleigh-Bénard convection is investigated with sulfur hexafluoride ($\mathrm{SF}_6$) in the vicinity of its critical point. In the supercritical domain, direct measurements of the heat flux $Q$ as a function of the temperature difference $\Delta T$ are consistent with the usual scaling laws of single-phase turbulent convection. Along the liquid-vapor coexistence curve, heat fluxes are dramatically enhanced by condensation and boiling. Optical measurements are performed to document the size and velocity of the bubbles. We report $Q(\Delta T, \epsilon)$ in both domains, with $\epsilon$ the dimensionless distance to the critical point.  Critical scaling laws are observed that can guide the development of theoretical models. In addition, this documents a system of diverging heat transfer coefficient, i.e., in which a significant heat flux can be achieved with an arbitrarily small temperature difference as $\epsilon \rightarrow 0$.
\end{abstract}

\maketitle
\paragraph*{Introduction.---} Engineering routinely exploits phase change to achieve high heat transfer rates: in particular, boiling and condensation can be found nearly everywhere in the industry, from micro-electronic cooling to large-scale nuclear reactors. Despite its ubiquity, relations between these heat fluxes and simple quantities such as temperature differences still consist of phenomenological correlations and semi-theoretical models, machine learning being now considered instead of theory-driven approaches to rationalize the various and often inconsistent datasets \cite{AI}. This lack of unified understanding should come as no surprise given the range of intricate phenomena involved (e.g., the entrapment of vapor in micro-cavities with moving interfaces surrounded by highly turbulent flows) and the reported dependency on fine experimental properties (e.g., surface roughness \cite{THEOFANOUS2002775, ALMASRI201724}). Experiments close to the critical point can help unveil some fundamental features of this complex dynamics by probing fluids in which thermodynamic quantities such as the latent heat can be continuously varied with a given setup of fixed surface roughness. The seminal work of reference  \cite{PhysRevLett.102.124501} evidenced a dramatic increase of the heat flux in Rayleigh-Bénard convection along the liquid-vapor coexistence curve as the system approaches the critical point. Since the latent heat vanishes in this limit, the authors attributed this feature to a divergence of the bubble nucleation rate. In this letter, we confirm this heat flux enhancement and investigate additional properties. In particular we report the direct measurement of the heat flux as a function of the external temperature difference and of the mean temperature both in the single-phase supercritical domain and along the liquid-vapor coexistence curve. In addition, optical measurements reveal that the mean radius of the bubbles obeys a non-trivial scaling law.

\begin{figure}[htb]
\includegraphics[width=1\linewidth]{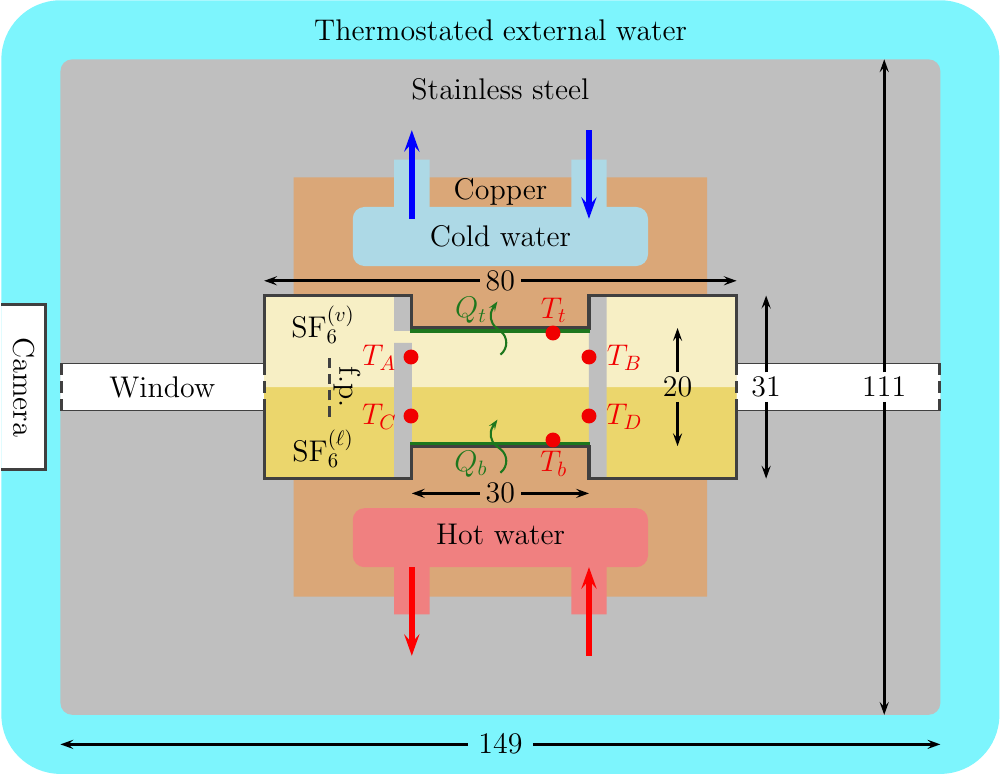}
\caption{\label{fig:epsart} Experimental setup (all dimensions are in mm). The focal plane (f.p.) is sketched as a vertical dashed line.}
\end{figure}

\paragraph*{Experimental setup.---}
The experimental setup is sketched in Fig. \ref{fig:epsart} and consists of a cylindrical container of inner radius $80~\mathrm{mm}$ and height $31~\mathrm{mm}$ filled with $\mathrm{SF}_6$ of purity $>99.7\%$. The total density $\rho$ has been set close to the critical density $\rho_\mathrm{c}$ so that in the vicinity of the critical point the liquid and vapor phases have similar volumes (they differ by less than 16\% in the data reported here). This choice was partly motivated by the previous investigation of Ref. \cite{PhysRevLett.102.124501} in which the heat flux is maximal for a volume fraction of liquid of about $1/2$. More precisely, comparison of our experimental measurements of the equation of state in the supercritical domain with the data from Ref. \cite{Biswas1984} yields $\rho = (740 \pm 2)~\mathrm{kg}\cdot \mathrm{m}^{-3}$ (reported values of $\rho_\mathrm{c}$ range between $736~\mathrm{kg}\cdot\mathrm{m}^{-3}$ and $742~\mathrm{kg}\cdot\mathrm{m}^{-3}$ \cite{Balzarini1974, WAGNER1992}). 

We thereafter focus on a subsystem of height $h = 20~\mathrm{mm}$ and rectangular cross section of size $L\times L$ with $L=30~\mathrm{mm}$, centered in the middle of the domain and bounded by $3~\mathrm{mm}$-thick PMMA walls. It is connected with the rest of the fluid by a small aperture at the top. Tracking the position of the interface when it exists indicates that the mean density in this subsystem can be considered constant within 1\%: the critical point is then reached by tuning the temperature of the boundaries to the critical one and we have no direct control on the mean pressure. Phase diagrams are reported as supplemental material and evidence the contrast with the experimental device of Ref. \cite{PhysRevLett.102.124501} in which the both the pressure and temperature were fixed (the mean density being not fixed). We shall see that the same qualitative behavior is observed in both devices.

The top and bottom temperatures and heat fluxes per unit surface, denoted as $T_t$, $Q_t$, $T_b$ and $Q_b$, are measured with Omega 44033 thermistors and Omega HFS-5 coupled with Keysight 34420A nanovoltmeters. Respectively hot and cold water provided by Lauda thermostatted baths flows through the bottom and top copper walls, generating a mean temperature difference $\Delta T = \langle T_b - T_t \rangle$ across the fluid ($\langle \cdot \rangle$ denotes a time average in a stationary state and we also define $T_m = \langle T_t + T_b \rangle /2$). This temperature difference $T_b-T_t$ is subject to typical fluctuations of $1~\mathrm{mK}$. The temperature of the water surrounding the cell, monitored with a Lauda thermostatted bath, is finely tuned to enforce $\langle Q_t \rangle = \langle Q_b \rangle \equiv Q$. Temperature measurements with Omega 44033 thermistors at $5$ and $15$-mm from the bottom heat flux sensor do not evidence large scale flows ($\langle T_A \rangle = \langle T_B \rangle$ and $\langle T_C \rangle = \langle T_D \rangle$, see Fig. \ref{fig:epsart}). Finally, a Basler acA2500-60uc camera equipped with a 70mm Sigma Macro lens is used to record $5~\mathrm{s}$-long movies of the two-phases dynamics with 200 frames per second. The focal plane is represented in Fig. \ref{fig:epsart}.

\begin{figure}[t]
\includegraphics[width=1\linewidth]{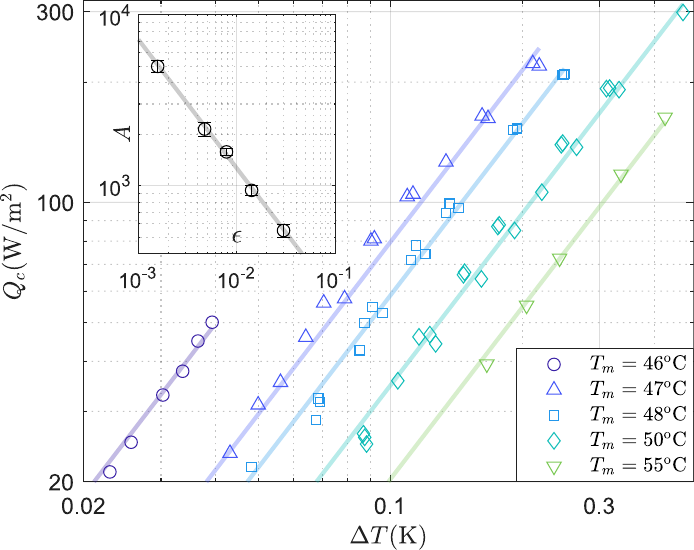}%
\caption{\label{fig:Q1p} Mean heat fluxes in the supercritical regime fitted by $Q_c(\Delta T, \epsilon) = A(\epsilon) \Delta T^{1.43}$. The size of the symbols is comparable to the standard deviation of the time series. Inset: $A(\epsilon)$ as a function of $\epsilon$ follows a critical law of exponent $-0.75 \pm 0.08$. These data correspond to Rayleigh numbers $Ra \in [2\times 10^9, 2\times 10^{11}]$ and Prandtl numbers $Pr\in [5, 60]$.}
\end{figure}

\paragraph*{Parameters and dimensionless numbers.---} Several physical properties undergo drastic changes as the mean temperature $T_m$ approaches the critical temperature $T_\mathrm{c} = 318.7~\mathrm{K}$, being proportional to some power of the dimensionless distance to the critical point ${\epsilon = \vert T_m - T_c \vert / T_c}$. In what follows, we denote with a subscript $c$ (respectively $\ell$ and $v$) the quantities measured in the supercritical domain along the critical isochore $\rho=\rho_c$ (resp. for liquid and vapor along the coexistence curve). The thermal diffusivity $\kappa$ as well as the difference of density between liquid and vapor $\rho_\ell - \rho_v$, the surface tension $\sigma_{\ell v}$, the latent heat $L_{\ell v}$ and the capillary length $\ell_{\ell v} = 2\pi \sqrt{\sigma_{\ell v}/(g(\rho_\ell - \rho_v))}$ vanish as $\epsilon \rightarrow 0$, with $g$ the gravitational acceleration (see, e.g., Ref. \cite{Jany_1990}). 
The kinematic viscosities of all phases $\nu$ are essentially constant in the range of temperature considered here. The isobaric expansion coefficient $\alpha$, the specific heat $c_p$ and the thermal conductivity 
$\lambda$ diverge as $\epsilon \rightarrow 0$. References \cite{Jany_1987, Lecoutre2009, Wu1973, SF6_Rathjen1980, 10.1063/1.2716004} are used to numerically evaluate these quantities. In the supercritical domain, we define the Rayleigh $Ra=g h^3\alpha_c \Delta T/(\nu_c \kappa_c)$, Prandtl $Pr = \nu_c / \kappa_c$ and Nusselt $Nu=Q_c/(\lambda_c \Delta T/h)$ numbers. Along the coexistence curve, the Jakob number $Ja= (\rho_\ell c_{p,\ell} \Delta T)/(2 \rho_v L_{\ell v})$ corresponds to the variation of internal energy between the bottom hot wall temperature $T_b = T_m +\Delta T/2$ at which boiling occurs and the mean bulk temperature $T_m$, normalized by the latent heat.

\begin{figure}[t]
\includegraphics[width=1\linewidth]{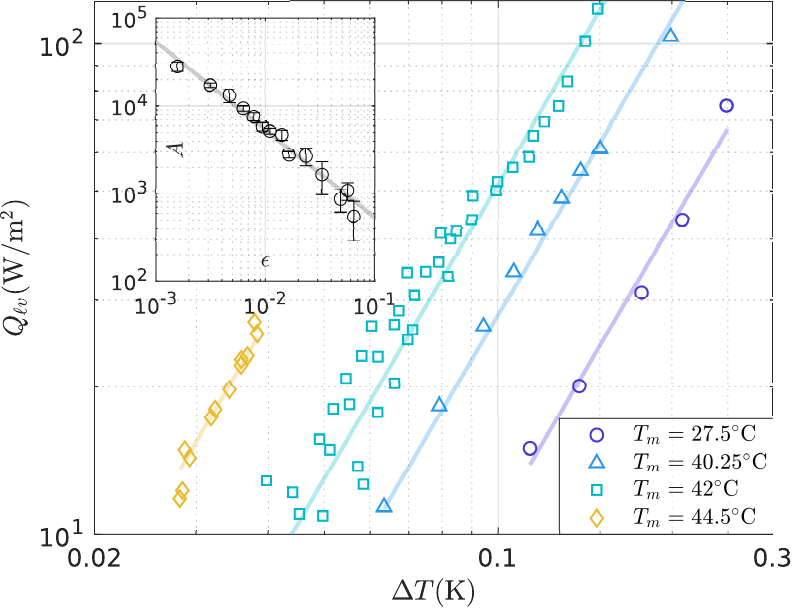}
\caption{\label{fig:Q2p} Mean heat fluxes along the liquid-vapor coexistence curve fitted by $Q_{\ell v}(\Delta T, \epsilon) = A(\epsilon) \Delta T^2$. The size of the symbols is comparable to the standard deviation of the time series. Inset: $A(\epsilon)$ as a function of $\epsilon$ follows a critical law of exponent $-1.00 \pm 0.09$. Note that, to improve readability, the main plot does not show data for all mean temperatures $T_m$ (the inset does). These data correspond to Jakob numbers $Ja \in [0.004, 0.025]$.}
\end{figure}

\paragraph*{Heat fluxes in the supercritical domain.---} The thermal properties of this system are first investigated for $T_b >T_t> T_\mathrm{c}$. Previous studies close to the critical point of helium \cite{Niemela2000, Chavanne2001,NIEMELA_SREENIVASAN_2006,  Roche_2010} and $\mathrm{SF}_6$ \cite{Steinberg1999}
have demonstrated that (i) extreme Rayleigh numbers $Ra$ can be achieved before non-Boussinesq effects become significant and (ii) that up to $Ra=O(10^{13})$ the turbulent heat flux scales as $Q_c \propto \Delta T^{1+\beta}$ (i.e.,  $Nu \propto Ra^\beta $) with $\beta \simeq 0.3$ usually compared to theoretical predictions of $2/7$ and $1/3$ \cite{Castaing_Gunaratne_Heslot_Kadanoff_Libchaber_Thomae_Wu_Zaleski_Zanetti_1989, Howard1966} (the behavior of the heat flux at even larger $Ra$ is still thoroughly discussed experimentally). Only small dependencies of $Nu$ on $Pr$ are reported. Note that, although accounting for the adiabatic temperature difference $\Delta T_\mathrm{ad}$ is crucial to capture the onset of convection close to the critical point \cite{Kogan1999}, this correction can be neglected here given the comparatively much larger temperature differences considered ($\Delta T_\mathrm{ad} < 2~\mathrm{mK}$).

Our measurements reveal that the heat flux scales as $Q_c(\Delta T, \epsilon) = A_c \epsilon^{-0.75\pm 0.08}\Delta T^{1+\beta} $ with $\beta = 0.43 \pm 0.13$ and $A_c=39~\mathrm{W}\cdot \mathrm{m}^{-2}$, see Fig. \ref{fig:Q1p}. The power-law scaling with $\epsilon$ confirms the expectation of a critical scaling behavior, and the negative exponent indicates that heat transfer coefficient $Q_c/\Delta T$ becomes extremely large as the system approaches the critical point. Using dimensionless numbers, the compensated Nusselt number $Nu Ra^{-\beta} $ is found almost constant and independent of $Pr$ ($Nu Ra^{-\beta} \in [0.03,0.04]$ for $Pr \in[5,60]$). The obtained value of $\beta$ is slightly larger than the ones usually reported but we stress that, given the experimental uncertainty, it remains compatible with previous observations. Indeed this experiment is not designed to measure precisely this exponent which would require larger variations of the Rayleigh numbers (see, e.g., Ref. \cite{Niemela2000}).  The temperature is homogeneous in the bulk ($\langle T_A \rangle =\langle T_B \rangle =\langle T_C \rangle =\langle T_D \rangle$) and does not significantly differ from $T_m$. These first results serve as a procedure to confirm the adequacy of our experimental setup for the study of heat fluxes close to the critical point, and shall be used for comparison with the two-phase regime.

\begin{figure}[t]
\includegraphics[width=1\linewidth]{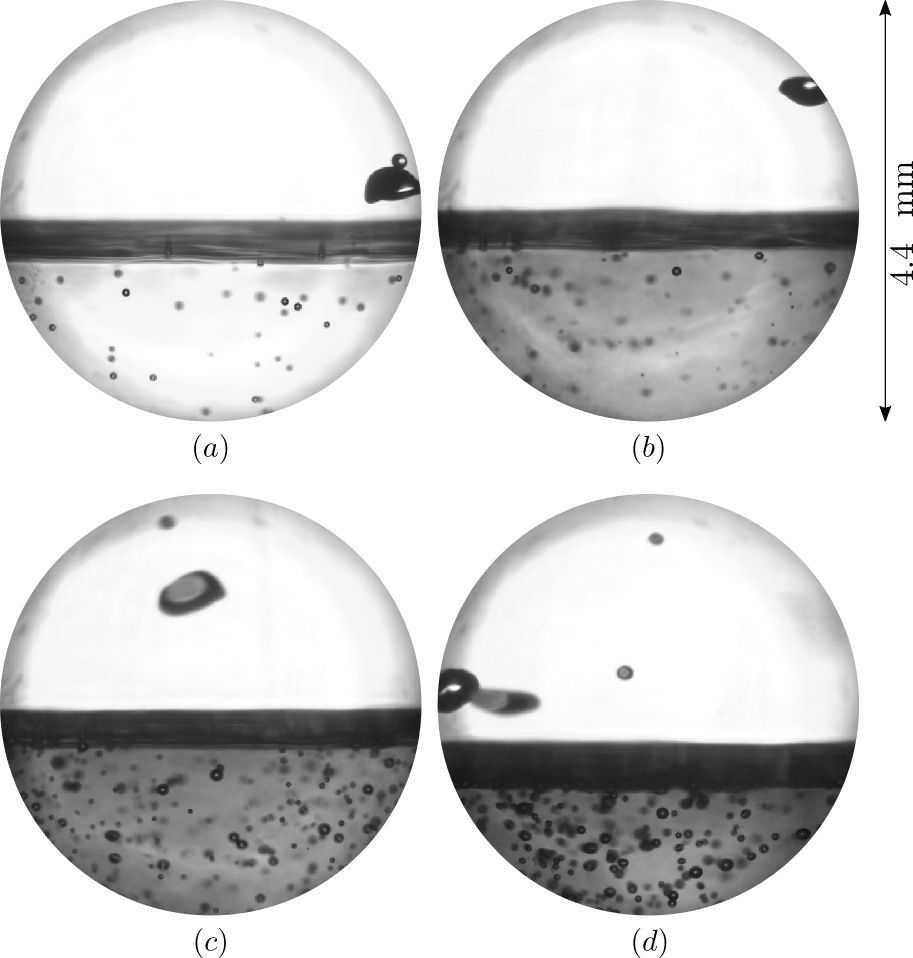}
\caption{\label{fig:snapshots} Typical snapshots at a fixed $T_m=40.25^\mathrm{o}\mathrm{C}$ for increasing $\Delta T$. A lot of rising vapor bubbles are observed in the liquid phase and a few liquid droplets falling in the vapor phase. $(a)$ $\Delta T = 0.064~\mathrm{K}$, $Q_{\ell v}= 11~\mathrm{W}/\mathrm{m}^2$, $(b)$ $\Delta T = 0.12~\mathrm{K}$, $Q_{\ell v}= 42~\mathrm{W}/\mathrm{m}^2$, $(c)$ $\Delta T = 0.15~\mathrm{K}$, $Q_{\ell v}= 61~\mathrm{W}/\mathrm{m}^2$ and $(d)$ $\Delta T = 0.20~\mathrm{K}$, $Q_{\ell v}= 103~\mathrm{W}/\mathrm{m}^2$.   The capillary length at that temperature is $\ell_{\ell v} = 1.3~\mathrm{mm}$. The corresponding movies are reported as supplemental material.}
\end{figure} 

\paragraph*{Heat fluxes along the liquid-gas coexistence curve.---} We now investigate heat fluxes as the system approaches the critical point from below, i.e., $T_c > T_b > T_t $. The fluid is now constituted of both liquid and vapor phases, with boiling occurring at the bottom (hot) boundary and condensation at the top (cold) one. As reported in Fig. \ref{fig:Q2p}, heat fluxes evidence different scaling laws than in the supercritical domain. More specifically, we find that the heat flux $Q_{\ell v}(\Delta T, \epsilon) = A_{\ell v} \epsilon^{-1.00 \pm 0.09}\Delta T^{2.00 \pm 0.13} $ with $A_{\ell v}=53~\mathrm{W}\cdot \mathrm{m}^{-2}$. Similar to the previous investigation of Ref. \cite{PhysRevLett.102.124501}, a divergence of the heat flux as $\epsilon \rightarrow 0$ is observed. However, we are able with the present experiment to document the exponents of this divergence and to demonstrate that it is more pronounced than in the supercritical domain  ($Q_{\ell v}/Q_c= 1.4 \epsilon^{-0.24}  \Delta T^{0.57}$). 
A wide range of correlation laws $Q_{\ell v} \propto \Delta T^m$ can be found in the literature for nucleate pool-boiling, with $m\in[1,4]$ and substantial discrepancies between heaters of different roughness \cite{Dhir, THEOFANOUS2002775, ALMASRI201724, SMITH196911, VanSciver:1444601}. The reported scaling $Q_{\ell v} \propto \epsilon^{-1.00 \pm 0.09} \Delta T^{2.00 \pm 0.13}$ can be used to provide a constraint for future models, given that most thermodynamic properties of $\mathrm{SF}_6$ are well documented as $\epsilon \rightarrow 0$.

Temperature is homogeneous in both phases ($\langle T_A\rangle = \langle T_B \rangle$ and $\langle T_C\rangle = \langle T_D\rangle$) but is systematically larger in the vapor than in the liquid. This inversion, which makes the vertical temperature profile non-monotonic, was previously reported in Ref. \cite{URBAN2013} and ascribed to heat fluxes being mainly driven by cold falling droplets and hot rising bubbles not in thermal equilibrium with their surrounding phases. While Ref. \cite{URBAN2013} only investigates transient two-phase convection, our results confirm that this inversion persists in a steady-state.
\begin{figure}[t]
\includegraphics[width=1\linewidth]{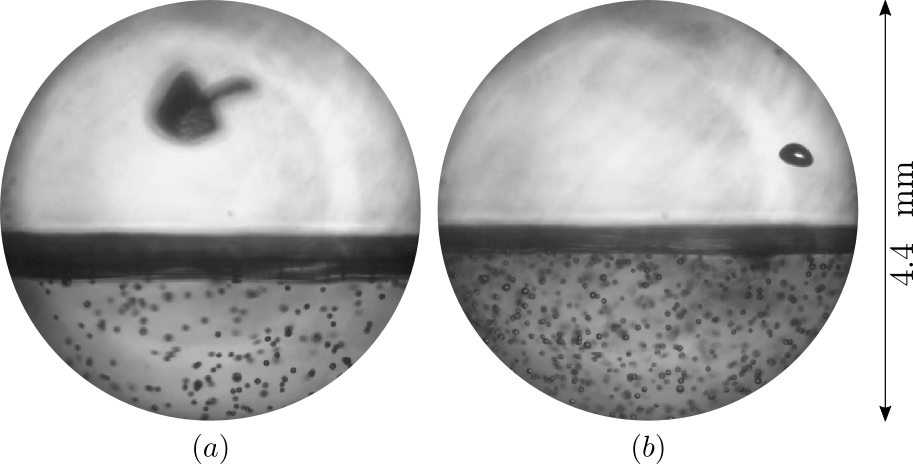}
\caption{\label{fig:snapshots2} Typical snapshots for various $T_m$ at similar ${Q_{\ell v} \sim 17~\mathrm{W}\cdot \mathrm{m}^{-2}}$ $(a)$ $T_m=35^\mathrm{o}\mathrm{C}$, $\Delta T = 0.08~\mathrm{K}$, $\ell_{\ell v} = 1.8~\mathrm{mm}$. $(b)$ $T_m=44^\mathrm{o}\mathrm{C}$, $\Delta T = 0.04~\mathrm{K}$, $\ell_{\ell v} = 0.7~\mathrm{mm}$.}
\end{figure}

\paragraph*{Bubble dynamics.---} To get a further insight into the dynamics leading to these enhanced heat fluxes, the movies recorded in the liquid-gas coexistence domain are analyzed. They evidence both rising bubbles from nucleate pool-boiling in the liquid phase and falling drops resulting from condensation on the top cold boundary. A few qualitative features can be directly drawn from the observation of snapshots. For fixed temperature $T_m$, Fig. \ref{fig:snapshots}  shows that both the typical radius and number of bubbles increase with $\Delta T$.  We also note that although the capillary length $\ell_{\ell v}$ relates to the wavenumber of the most unstable mode $k_{RT}$ of the inviscid Rayleigh Taylor instability ($2\pi/k_{RT} = \sqrt{3} \ell_{\ell v}$), it compares neither to the sizes of the droplet nor to the ones of the bubbles. This feature is evident in Fig. \ref{fig:snapshots2}, in which no drastic diminution of the size of the bubbles is observed, although the capillary length $\ell_{\ell v}$ drops by a factor of $2.5$.

\begin{figure}[t]
\includegraphics[width=1\linewidth]{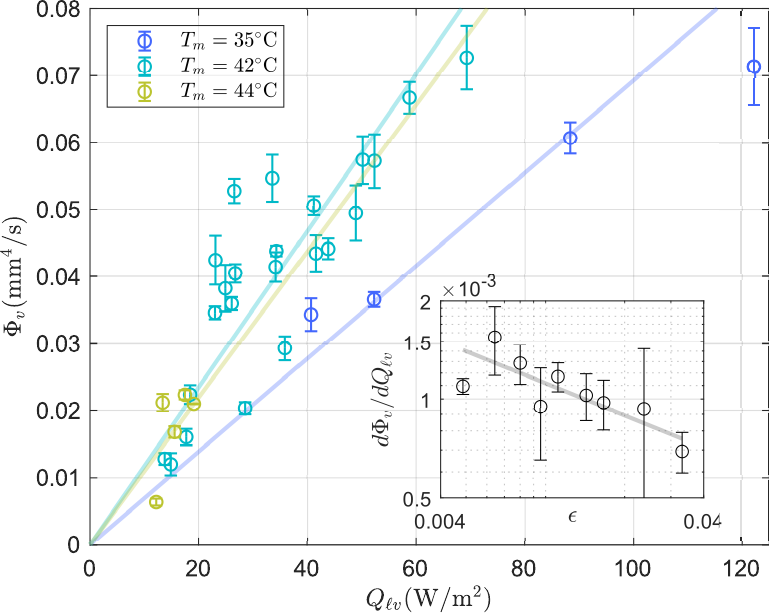}
\caption{\label{fig:V_vs_Q} Mean volumetric flux estimated from the movies as a function of the heat flux. Each set of data obtained at a given mean temperature $T_m$ is fitted by a straight line whose slope is reported in the inset as a function of the distance to the critical point $\epsilon$.   A power law of slope $\epsilon^{-0.325}$, proportional to the inverse latent heat $L_{\ell v}$ is also shown. Note that, to improve readability, the main plot does not show data for all mean temperatures $T_m$ (the inset does).}
\end{figure}

A quantitative analysis of the bubbles properties is now performed by measuring their position and radius $r_b$ with the two-stage circular Hough transform \cite{Davies}. When only a few isolated bubbles rise ($Q_{\ell v} < 20~\mathrm{W}\cdot \mathrm{m}^{-2}$), they can be tracked and their vertical velocity $v_b$ can in addition be measured. Given that the bubble Reynolds numbers $v_b r_b/\nu_\ell \in [3, 37]$,  $v_b$ results from a balance between buoyancy and drag and we checked that it can be accurately estimated using the rigid sphere approximation of Ref. \cite{Cliff} for the drag coefficient. To analyze more complex states in which a swarm of $N_b \gg 1$ bubbles continuously rises, a mean rise velocity $\langle v \rangle$ is measured by finding the largest cross correlation between successive pictures. A mean volumetric flux $\Phi_v = \langle v \rangle N_b \langle r_{b}^3 \rangle $ can then be estimated, defined as the product of this mean rise velocity with the mean vapor volume on the snapshots area. Given that the procedure to find the bubbles is found accurate only for radius larger or equal than three pixels ($20~\mu \mathrm{m}$), we thereafter restrict ourselves to datasets for which the mean radius is found larger than five pixels ($33~\mu \mathrm{m}$), in order for the small neglected bubbles not to significantly alter the results. Note that $\Phi_v$ is a local measurement outside the domain of interest (see Fig. \ref{fig:epsart}) that depends on the depth of field and therefore only roughly approximates the total volume of bubbles being generated by unit time. Nevertheless, it is found approximately proportional to $Q_{\ell v}$, see Fig. \ref{fig:V_vs_Q}. In addition, as evidenced by the inset of Fig. \ref{fig:V_vs_Q}, the slope $\mathrm{d}\Phi_v /\mathrm{d}Q_{\ell v}$ is consistent with a divergence as $\epsilon^{-0.325}$: since the latent heat vanishes as $L_{\ell v} \propto \epsilon^{0.325}$, this confirms that the heat flux in this system is dominated by phase changes ($Q_{\ell v} \propto L_{\ell v} \Phi_v$). Coupled with the direct measurement $Q_{\ell v} \propto \epsilon^{-1}\Delta T^2$, it yields $\Phi_v\propto \epsilon^{-1.3} \Delta T^2$. A detailed analysis of each term of $\Phi_v$ ($\langle v \rangle$ and $N_b$ are reported as supplemental material, $\langle r_b^3 \rangle$ is discussed below) shows that the heat flux diverges close to the critical point as a consequence of a rapid increase of the number and size of the nucleated bubbles.

\begin{figure}[t]
\includegraphics[width=1\linewidth]{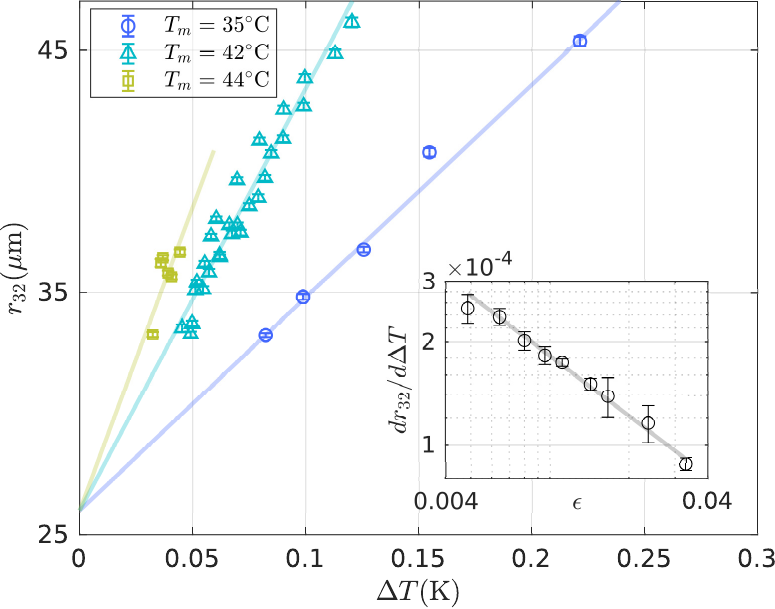}
\caption{\label{fig:mean_R} Sauter mean radius $r_{32}(\epsilon, \Delta T)$ as a function of the temperature difference $\Delta T$. A fit by $a(\epsilon) \Delta T + 26 \mu \mathrm{m}$ is performed and the inset reports $a(\epsilon)$ along with its critical law fit of exponent $-0.58 $. Note that, to improve readability, the main plot does not show data for all mean temperatures $T_m$ (the inset does).}
\end{figure}

Finally, we discuss the radius of the bubbles as a function of the temperature difference $\Delta T$ for various mean temperatures $T_m$. The Sauter mean radius ${r_{32}= \langle r_b^3\rangle / \langle r_b^2 \rangle}$ is investigated to minimize the bias resulting from small bubbles not being optically detected, but very similar observations are reached based on the mean radius $\langle r_b \rangle$. As shown in Fig. \ref{fig:mean_R}, our data can be modeled by $r_{32}(\epsilon, \Delta T)= r_0 + r_1 \epsilon^{-0.58 \pm 0.05} \Delta T$, with $r_0$ and $r_1$ two constants. The range of Jakob numbers $Ja \ll 1$ ({$Ja \in [0.004, 0.025]$}) that has been explored rationalizes the experimental observation that the bubbles do not grow or shrink as they rise, and we therefore access their radius at nucleation. A large number of models have been proposed for the mean radius of bubbles in nucleate pool boiling as they leave the bottom boundary, see, e.g., Table 3 of Ref. \cite{MAHMOUD2021101024}. They essentially consider that it depends on a combination of (i) the contact angle $\theta$, (ii) the capillary length $\ell_{\ell v}$, and (iii) the Jakob number $Ja$. The roughness length scale $r_r$ also plays a crucial role in nucleation. Close to the critical point, perfect wetting occurs \cite{10.1063/1.434402} and $\theta$ can be disregarded in the present experiment. For the range of finite $\epsilon$ that is explored (for which the theoretical critical scaling laws derived in the limit $\epsilon \rightarrow 0$ do not systematically hold), we can approximate $r_r \propto \epsilon^0$, $\ell_{\ell v}\propto \epsilon^{0.5}$ and $Ja \propto \epsilon^{-1.1} \Delta T $. Our dataset could therefore be interpreted as $r_{32} = r_r + c\times Ja \times \ell_{\ell v}$, with $c$ a dimensionless constant and $r_r=26~\mu\mathrm{m}$. It corresponds to the scaling law obtained by \cite{Cole67} ($r_{32} \propto Ja \times \ell_{\ell v}$) corrected by a finite roughness $r_r$ that becomes significant since both $Ja \ll 1$ and $\ell_{\ell v}$ remain small ($\ell_{\ell v} \in [0.6, 2]~\mathrm{mm}$). In contrast, a typical boiling experiment in water corresponds to $Ja = O(10)$ and $\ell_{\ell v} = 17~\mathrm{mm}$.

\paragraph*{Conclusion.---} A Rayleigh-Bénard device filled with fluid at critical density allows the comparison of monophasic convection (supercritical fluid domain) with biphasic convection (liquid-vapor coexistence range). The major difference between these is the possibility of phase change driven heat fluxes, involving the nucleation of bubbles and the condensation of drops. The present experiment evidences that, although in both domains the heat transfer coefficient $Q/\Delta T$ diverges as $\epsilon \rightarrow 0$, this divergence is much more pronounced in the biphasic domain. In particular, the usual single phase turbulent scaling law for the heat flux that describes our results in the supercritical domain does not hold in the liquid-vapor regime and a new scaling is observed. We have shown that this new regime results from heat fluxes being driven by phase change, with a number of nucleated bubbles that rapidly increases as the critical point is approached. This also generates non-monotonic vertical temperature profiles. Our reported measurements of $Q_{\ell v}(\Delta T, \epsilon)$ both as a function of $\Delta T$ and $\epsilon$ can help develop models to capture this complex dynamics. Moreover, direct measurements of the radius of rising  bubbles evidence a non-trivial scaling law that depends on both the surface roughness, the capillary length and the Jakob number. In particular, nucleation close to the critical point is found associated with very small bubbles, whose size compares to the surface roughness that therefore becomes a key parameter.

\paragraph*{Acknowledgments.---}
 This work was funded by the French National Research Agency (ANR LASCATURB, No. ANR-23-CE30-0043) and by CNES (Centre National d'\'Etudes Spatiales).

\providecommand{\noopsort}[1]{}\providecommand{\singleletter}[1]{#1}%

\end{document}